\documentclass[conference]{IEEEtran}

\usepackage{amsmath, amsfonts} 
\usepackage{amsthm}
\usepackage[cmintegrals]{newtxmath}
\usepackage{graphicx}
\usepackage{multirow}
\usepackage[hidelinks]{hyperref} 

\newcommand{\overbar}[1]{\mkern 1.5mu\overline{\mkern-1.5mu#1\mkern-1.5mu}\mkern 1.5mu}

\newtheorem{newdef}{Definition}
\newtheorem{theorem}{Theorem}

\begin{document}

\title{The Impact of Partial Packet Recovery on the Inherent Secrecy of Random Linear Coding}

\author{\IEEEauthorblockN{Ioannis~Chatzigeorgiou}
\IEEEauthorblockA{School of Computing and Communications\\
Lancaster University, United Kingdom\\
Email: i.chatzigeorgiou@lancaster.ac.uk}}

\maketitle
\begin{abstract}
This paper considers a source, which employs random linear coding (RLC) to encode a message, a legitimate destination, which can recover the message if it gathers a sufficient number of coded packets, and an eavesdropper. The probability of the eavesdropper accumulating enough coded packets to recover the message, known as the intercept probability, has been studied in the literature. In our work, the eavesdropper does not abandon its efforts to obtain the source message if RLC decoding has been unsuccessful; instead, it employs partial packet recovery (PPR) offline in an effort to repair erroneously received coded packets before it attempts RLC decoding again. Results show that PPR-assisted RLC decoding marginally increases the intercept probability, compared to RLC decoding, when the channel conditions are good. However, as the channel conditions deteriorate, PPR-assisted RLC decoding significantly improves the chances of the eavesdropper recovering the source message, even if the eavesdropper experiences similar or worse channel conditions than the destination.
\vspace{2pt}
\end{abstract}



\section{Introduction}
\label{sec.Introduction}

Random linear coding (RLC), which encompasses fountain coding and network coding,
generates coded packets that are random linear combinations of input data packets. RLC has the potential to increase throughput and improve robustness~\cite{Bassoli2013} but has also been considered for information-theoretic secure communications because eavesdroppers cannot necessarily retrieve source data from a subset of received coded packets. Ning and Yeung \cite{Cai2002} first formulated the concept of secure network coding, while Adeli and Liu \cite{Adeli2013} studied probabilistic weak security for linear network coding. Niu \textit{et al.} \cite{Niu2014} computed the intercept probability of RLC for wireless broadcast applications, that is, the probability that an eavesdropper will accumulate the required number of coded packets for the reconstruction of the source data. The analysis in \cite{Niu2014} was later revisited and refined by Khan~\textit{et al.} \cite{Khan2015b}. Since then, the intercept probability has been analyzed in different settings; for example, layered RLC was studied in \cite{Karim2017}, sparse RLC was investigated in \cite{Tassi2019a} and applications of secure RLC-encoded data transfer were explored in \cite{Tassi2019b,Sun2020}.

RLC decoding at a legitimate destination or an eavesdropper discards received coded packets that have been corrupted by errors and attempts to reconstruct the source packets from correctly received coded packets. In this paper, we assume that an eavesdropper does not discard corrupted coded packets but stores them and utilizes them offline, i.e., at a later instance, if RLC decoding has been unsuccessful. Partial packet recovery (PPR), proposed in~\cite{Mohammadi2016}, is employed to repair stored packets in polynomial time and improve the chances of the eavesdropper recovering the source packets. The motivation for this paper is to investigate RLC decoding combined with PPR from the perspective of an eavesdropper, study the impact of PPR-assisted RLC decoding on the intercept probability, contrast it with conventional RLC decoding and observe tradeoffs between secrecy and reliability.

The remainder of this paper has been organized as follows: Section~\ref{sec.System_Model} presents the system model, including definitions and notation. Section~\ref{sec.RLC_encoding} proposes full-rank nonsystematic RLC encoding at the source, while Section~\ref{sec.PPR_at_eavesdropper} studies PPR-assisted RLC decoding at an eavesdropper and briefly describes two methods for repairing corrupted packets. The chances that a legitimate destination or an eavesdropper will recover the source message are discussed in Section~\ref{sec.InterceptProb}. Results, which give insights into secrecy-reliability tradeoffs, are presented in Section~\ref{sec.Results} and key findings are summarized in Section~\ref{sec.Conclusions}.

\section{System Model}
\label{sec.System_Model}

\begin{figure}[t]
\centering
\includegraphics[width=0.97\columnwidth]{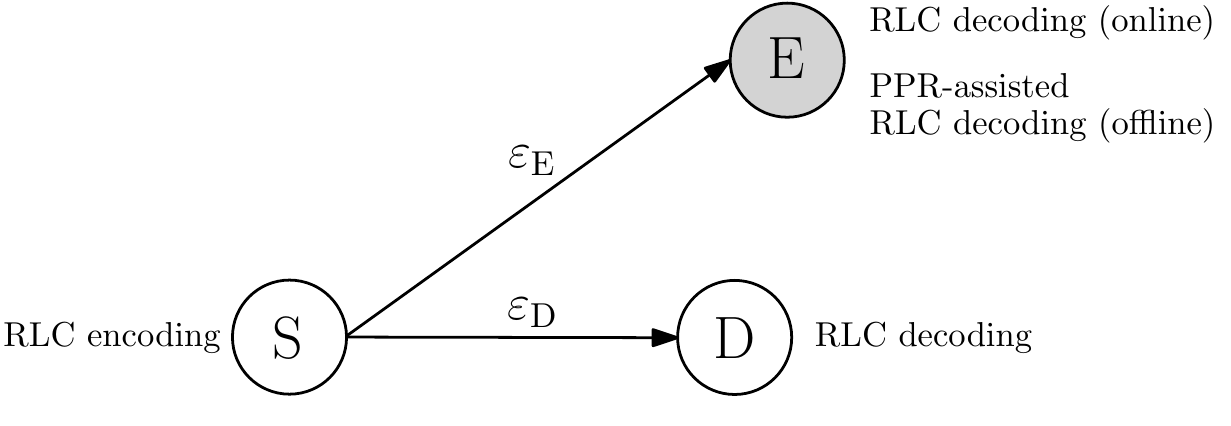}
\caption{Diagram of the system model, where $\varepsilon_\mathrm{D}$ and $\varepsilon_\mathrm{E}$ denote the packet error probabilities of the channels linking the source~$\mathrm{S}$ to the destination~$\mathrm{D}$ and the eavesdropper~$\mathrm{E}$. Both the destination and the eavesdropper employ RLC decoding to recover the source message. The eavesdropper resorts to PPR-assisted RLC decoding, if RLC decoding alone has been unsuccessful.}
\label{fig:system_model}
\end{figure}

We consider a network setup whereby a source $\mathrm{S}$ transmits a message to a legitimate destination $\mathrm{D}$ in the presence of an eavesdropper $\mathrm{E}$, as depicted in Fig.~\ref{fig:system_model}. Before initiating transmission, the source segments the message into $K$ packets. Each source packet has been modeled as a sequence of $L$ symbols from a finite field of $q$ elements, denoted by $\mathbb{F}_{q}$, where $q$ is a prime power. The $K$ source packets of length $L$ can be expressed as a matrix $\mathbf{U}\in\mathbb{F}^{K\times L}_{q}$, where $\mathbb{F}^{K\times L}_{q}$ denotes the set of all $K\times L$ matrices over $\mathbb{F}_{q}$. RLC~\cite{Ho2006} is used to encode the $K$ source packets into $N\geq K$ coded packets, each of length $L$ symbols. The $N$ coded packets can also be expressed in matrix form as $\mathbf{X}\in\mathbb{F}^{N\times L}_{q}$. The relationship between matrices $\mathbf{X}$ and $\mathbf{U}$ is:
\begin{equation}
\label{eq.RLNC_matrix}
\mathbf{X} = \mathbf{G}\,\mathbf{U},
\end{equation}
where $\mathbf{G}\in\mathbb{F}^{N\times K}_{q}$ is known as the \textit{generator matrix}. Typically, the entries of $\mathbf{G}$ are chosen uniformly at random from $\mathbb{F}_{q}$. As explained in \cite{Mohammadi2016}, a seed can be used to initialize the pseudo-random number generator that outputs the entries of $\mathbf{G}$. Given that the value of the seed can be conveyed in the headers of the coded packets, we assume that both the destination and the eavesdropper have knowledge of $\mathbf{G}$. Note that the eavesdropper could be a legitimate destination in a different occasion -- hence, it is equipped with the same pseudo-random number generator -- but is not authorized to receive the private source message in this occasion. The link connecting the source to the destination has been modeled as a memoryless channel characterized by packet error probability $\varepsilon_\mathrm{D}$ or, equivalently, by bit error probability $1-(1-\varepsilon_\mathrm{D})^{1/L}$. Similarly, the link connecting the source to the eavesdropper is characterized by packet error probability $\varepsilon_\mathrm{E}$. We assume that the eavesdropper experiences, on average, similar or worse channel conditions than the destination, that is, $\varepsilon_\mathrm{E}\geq\varepsilon_\mathrm{D}$.

Let $N_\mathrm{D}$ be the number of coded packets that have been successfully received by the destination, and $\mathcal{D}$ be the ordered set of the row indices of $\mathbf{X}$ that correspond to correctly received coded packets, i.e., $\vert \mathcal{D}\vert=N_\mathrm{D}\leq\! N$. The destination constructs matrix $\mathbf{X}_\mathcal{D}$ from the $N_\mathrm{D}$ correctly received coded packets and matrix $\mathbf{G}_\mathcal{D}$ from $N_\mathrm{D}$ rows of $\mathbf{G}$ with indices in $\mathcal{D}$. The source message, represented by matrix $\mathbf{U}$, can be obtained from $\mathbf{X}_\mathcal{D} = \mathbf{G}_\mathcal{D}\,\mathbf{U}$ if the rank of $\mathbf{G}_\mathcal{D}$ is $K$. In that case, the relationship $\mathbf{X}_\mathcal{D} = \mathbf{G}_\mathcal{D}\,\mathbf{U}$ can be seen as a system of $N_\mathrm{D}\geq K$ linear equations, which can be reduced to a system of $K$ linearly independent equations with $K$ unknowns, i.e., source packets. This $K\times K$ system can be solved, e.g., using Gaussian elimination, and return a unique solution for $\mathbf{U}$. Similarly, the overhearing eavesdropper collects $N_\mathrm{E}$ coded packets, builds the ordered set $\mathcal{E}$ of row indices, constructs matrices $\mathbf{X}_\mathcal{E}$ and $\mathbf{G}_\mathcal{E}$, and attempts to obtain $\mathbf{U}$ from $\mathbf{X}_\mathcal{E} = \mathbf{G}_\mathcal{E}\,\mathbf{U}$.

\textit{Khan et al.} considered this system in \cite{Khan2015b} and derived closed-form expressions for the probability that the eavesdropper will gather enough coded
packets to recover the source message. The authors also quantified the gain in secrecy when the source broadcasts up to $N_\mathrm{max}$ coded packets but ceases transmission as soon as the destination sends a notification over a perfect feedback channel confirming receipt of the source message. In this paper, the eavesdropper attempts to decode the $N_\mathrm{E}$ correctly received coded packets concurrently with the legitimate destination, as in \cite{Khan2015b}, but does not discard the $N-N_\mathrm{E}$ erroneously received coded packets. Instead, the eavesdropper employs PPR \cite{Mohammadi2016} offline in an effort to repair received coded packets that have been corrupted by errors and improve the chances of the RLC decoder recovering the source message. Section~\ref{sec.RLC_encoding} discusses RLC encoding at the source, while Section~\ref{sec.PPR_at_eavesdropper} adapts the PPR process presented in \cite{Mohammadi2016} to the system model under consideration and describes PPR-assisted RLC decoding at the eavesdropper.


\section{RLC Encoding at the Source}
\label{sec.RLC_encoding}

RLC encoding is typically \textit{non-systematic}, that is, the $K$ source packets are encoded into $N$ packets, which are random linear combinations of the source packets, as described in Section \ref{sec.System_Model}. The probability that the legitimate destination will successfully recover the $K$ source packets from $N_\textrm{D}\leq N$ received coded packets is given by \cite{Trullols-Cruces2011}:
\begin{equation}
\label{eq:cond_decoding_prob_nonsys}
P_\mathrm{ns}(N_\textrm{D},K) = \prod_{i=0}^{K-1}\left[1-q^{-\left(N_\textrm{D}-i\right)}\right].
\end{equation}

RLC encoding can also be \textit{systematic}, in which case the first $K$ of the $N$ transmitted packets are identical to the $K$ source packets while the remaining $N-K$ packets are random linear combinations of the source packets. In systematic RLC, the generator matrix $\mathbf{G}$ can be expressed in standard form:
\begin{equation}
\label{eq.sys_generator_matrix}
\mathbf{G}=
\left[\!\begin{array}{c} \mathbf{I}_K\\ \mathbf{P}\end{array}\!\right],
\end{equation}
where $\mathbf{I}_K$ is the $K\times K$ identity matrix and $\mathbf{P}$ is a \mbox{$(N-K)\times K$} randomly-generated matrix with entries from $\mathbb{F}_q$. The probability of the destination recovering the $K$ source packets, upon receipt of $N_\mathrm{D}\leq N$ error-free packets, can be calculated from:
\begin{equation}
\label{eq:cond_decoding_prob_sys}
P_\mathrm{s}(N,N_\textrm{D},K)=%
\sum_{h=h_\mathrm{lim}}^{K}\!\!
\resizebox{0.106\textwidth}{!}{$
\frac{\displaystyle\binom{K}{h}%
\binom{N-K}{N_\mathrm{D}-h}}%
{\displaystyle\binom{N}{N_\mathrm{D}}}
$}
\,P_\mathrm{ns}(N_\textrm{D}-h,\,K-h),
\end{equation}
where $h_\mathrm{lim}=\max{(0,\,N_\mathrm{D}+K-N)}$~\cite{Jones15}.

In systematic RLC, the eavesdropper could receive some of the source packets and could thus gain direct access to parts of the source message even before it attempts to decode the collected packets. Non-systematic RLC offers the advantage of inherent secrecy but exhibits a lower probability of message recovery at the destination than systematic RLC, as proven in \cite{Jones15}. For example, let us assume that the destination successfully received the first $K$ transmitted packets. If systematic RLC has been used, the destination will recover the source message. If non-systematic RLC has been employed, recovery of the source message is not guaranteed, as the first $K$ transmitted packets may not be linearly independent. In that case, the destination will need more than $K$ packet transmissions at the risk of allowing the eavesdropper to collect more packets and possibly recover the source message.

In order to combine the inherent secrecy of non-systematic RLC with the reliability of systematic RLC, we could use non-systematic RLC constructions defined by a generator matrix of the form:
\begin{equation}
\label{eq.pfx_nonsys_generator_matrix}
\mathbf{G}=
\left[\!\begin{array}{c} \mathbf{G}_K\\ \mathbf{P}\end{array}\!\right].
\end{equation}
To obtain $\mathbf{G}$, the seed of the pseudo-random number generator should be chosen to ensure that the top $K$ randomly-generated rows of $\mathbf{G}$ are linearly independent, that is, $\mathrm{rank}(\mathbf{G}_K)=K$. Column-wise Gaussian elimination can transform $\mathbf{G}$ into standard form and convert the proposed \textit{full-rank non-systematic RLC} into an equivalent systematic RLC, i.e.,
\begin{equation}
\label{eq.pfx_nonsys_to_sys}
\mathbf{G}=
\left[\!\begin{array}{c} \mathbf{G}_K\\ \mathbf{P}\end{array}\!\right]%
\xrightarrow[\text{column operations}]{\text{elementary}}%
\mathbf{G}'=
\left[\!\begin{array}{c} \mathbf{I}_K\\ \mathbf{P}'\end{array}\!\right].
\end{equation}
Full-rank non-systematic RLC retains the inherent secrecy features of non-systematic RLC, while its equivalence to systematic RLC suggests that \eqref{eq:cond_decoding_prob_sys} provides the probability of the destination recovering the $K$ source packets when $N_\mathrm{D}\leq N$ coded packets have been received without errors. Expression \eqref{eq.pfx_nonsys_to_sys} will be used in the following section, which looks into how the eavesdropper attempts to retrieve the source message.


\section{Eavesdropping Enhanced by PPR}
\label{sec.PPR_at_eavesdropper}

We explained in Section~\ref{sec.System_Model} that $\mathcal{E}$ is the ordered set of $N_\mathrm{E}$ row indices of $\mathbf{X}$ that correspond to error-free coded packets received by the eavesdropper. The indices of the remaining $N-N_\mathrm{E}$ rows of $\mathbf{X}$, which identify erroneously received coded packets stored at the eavesdropper, form $\overbar{\mathcal{E}}=\{1,\ldots,N\}\backslash\mathcal{E}$. Based on $\mathcal{E}$, the eavesdropper constructs $\mathbf{X}_\mathcal{E}$ and $\mathbf{G}_\mathcal{E}$, and uses RLC decoding to solve $\mathbf{X}_\mathcal{E} = \mathbf{G}_\mathcal{E}\,\mathbf{U}$ for $\mathbf{U}$ and recover the source message. If $\mathrm{rank}(\mathbf{G}_\mathcal{E})=K$, the source message can be recovered. If $\mathrm{rank}(\mathbf{G}_\mathcal{E})<K$, the eavesdropper turns to set $\overbar{\mathcal{E}}$, and utilizes sparse recovery principles to repair erroneously received coded packets and complement RLC decoding.

More specifically, let $\mathbf{Y}$ be an erroneous copy of $\mathbf{X}$ that has been received by the eavesdropper, defined as:
\begin{equation}
\label{eq.PPR_Y}
	\mathbf{Y}=\mathbf{X}+\mathbf{E}.
\end{equation}
The \textit{error matrix} $\mathbf{E}$ contains non-zero elements in positions where errors have occurred and zero elements in the remaining positions. Note that the eavesdropper has knowledge of $\mathbf{Y}$ and $\mathbf{G}$ only. The $N\times(N-K)$ \textit{parity-check matrix} $\mathbf{H}$ can be derived from the $N\times K$ generator matrix $\mathbf{G}$, as follows:
\begin{equation}
\label{eq.fullranl_nonsys_parity_check_matrix}
\mathbf{H}=\left[\!\begin{array}{c} -\mathbf{P}'\;\vert\;\mathbf{I}_{N-K}\end{array}\!\right]^\top,
\end{equation}
so that:
\begin{equation}
\label{eq.PPR_zero_product}
\mathbf{H}^\top\,\mathbf{G}=\mathbf{0}.
\end{equation}
Matrix $\mathbf{P}'$ can be obtained from \eqref{eq.pfx_nonsys_to_sys}. Matrix negation in \eqref{eq.fullranl_nonsys_parity_check_matrix} is performed in $\mathbb{F}_q$, e.g., $-\mathbf{P}'=\mathbf{P}'$ in $\mathbb{F}_2$. Multiplication of~$\mathbf{Y}$ by $\mathbf{H}^\top$ produces the $(N-K)\times L$ \textit{syndrome matrix} $\mathbf{S}$, i.e., $\mathbf{S} = \mathbf{H}^\top \mathbf{Y}$. Using \eqref{eq.PPR_zero_product}, we find that the relationship between the syndrome matrix $\mathbf{S}$ and the error matrix $\mathbf{E}$ is:
\begin{equation}
\label{eq.syndrome_decoding_full}
\mathbf{S}  = \mathbf{H}^\top \mathbf{Y}
 = \mathbf{H}^\top (\mathbf{G}\mathbf{U}+\mathbf{E})
 = \mathbf{H}^\top \mathbf{E}.
\end{equation}
Given that our focus is on the $N-N_\mathrm{E}$ erroneously received coded packets, we use the set $\overbar{\mathcal{E}}$ to isolate the $N-N_\mathrm{E}$ of the $N$ rows of $\mathbf{Y}$, $\mathbf{X}$, $\mathbf{E}$ and $\mathbf{H}$, and construct $\mathbf{Y}_{\overbar{\mathcal{E}}}$, $\mathbf{X}_{\overbar{\mathcal{E}}}$, $\mathbf{E}_{\overbar{\mathcal{E}}}$ and $\mathbf{H}_{\overbar{\mathcal{E}}}$. As a result, expressions \eqref{eq.PPR_Y} and \eqref{eq.syndrome_decoding_full} change to:
\begin{equation}
\label{eq.PPR_Y_corrupted} \mathbf{Y}_{\overbar{\mathcal{E}}}=\mathbf{X}_{\overbar{\mathcal{E}}}+\mathbf{E}_{\overbar{\mathcal{E}}}
\end{equation}
and
\begin{equation}
\label{eq.syndrome_decoding_partial}
\mathbf{S} = \left(\mathbf{H}_{\overbar{\mathcal{E}}}\right)^{\!\top} \mathbf{E}_{\overbar{\mathcal{E}}},
\end{equation}
respectively. If the $j$-th column of $\mathbf{S}$ and $\mathbf{E}_{\overbar{\mathcal{E}}}$ is denoted by $[\mathbf{S}]_{*,j}$ and $[\mathbf{E}_{\overbar{\mathcal{E}}}]_{*,j}$, expression \eqref{eq.syndrome_decoding_partial} can be re-written as $L$ independent systems of $N-K$ linear equations with $N-N_\mathrm{E}$ unknowns per equation:
\begin{equation}
\label{eq.syndrome_decoding_partial_per_col}
\left[\mathbf{S}\right]_{*,j} = \left(\mathbf{H}_{\overbar{\mathcal{E}}}\right)^{\!\top} \left[\mathbf{E}_{\overbar{\mathcal{E}}}\right]_{*,j}\quad\text{for}\quad j=1,\ldots,L.
\end{equation}

Mohammadi~\textit{et al.}~\cite{Mohammadi2016} observed that erroneously received coded packets usually contain large error-free segments, thus $\mathbf{E}_{\overbar{\mathcal{E}}}$ is a \textit{sparse} matrix, that is, $\mathbf{E}_{\overbar{\mathcal{E}}}$ has more zero entries than non-zero entries. Based on this observation, the solution to \eqref{eq.syndrome_decoding_partial_per_col} can be formulated as an `$\ell_0$ minimization' problem:
\begin{IEEEeqnarray}{ll}
\label{eq.L0_minimization}
\left[\hat{\mathbf{E}}_{\overbar{\mathcal{E}}}\right]_{*,j}=\;  &\arg\min_{\mathbf{w}^{\top}}\;\lVert \mathbf{w} \rVert_0 \IEEEyesnumber\IEEEyessubnumber*\label{eq.obj_func}\\
&\text{subject to}\; \left(\mathbf{H}_{\overbar{\mathcal{E}}}\right)^{\!\top}\mathbf{w}^{\top}=\left[\mathbf{S}\right]_{*,j}
\label{eq.constraint}
\end{IEEEeqnarray}
where $\mathbf{w}\in\mathbb{F}^{N-N_\mathrm{E}}_q$ is a row vector that should satisfy constraint \eqref{eq.constraint} and have the minimum possible number of non-zero entries. To count the non-zero entries in $\mathbf{w}$, the $\ell_0$ norm is used, which is defined as $\lVert \mathbf{w} \rVert_0=\lvert w_1\rvert^0+\ldots+\lvert w_{N-N_\mathrm{E}}\rvert^0$ assuming that $0^0=0$ \cite{Donoho2001}.

Two PPR methods for the evaluation of $\mathbf{\hat{E}}_{\overbar{\mathcal{E}}}$ were proposed in \cite{Mohammadi2016}. The first approach, which is inspired by the compressed sensing (CS) literature, replaces the non-convex $\ell_0$ norm in \eqref{eq.obj_func} with the convex $\ell_1$ norm, solves the optimization problem over the set of real numbers, and rounds off the derived values to the nearest elements of $\mathbb{F}_q$. The second approach, referred to as \textit{syndrome decoding}, considers \eqref{eq.L0_minimization} and initiates an exhaustive search for a candidate solution; the sparsity of the row vector $\mathbf{w}$ is gradually reduced, i.e., the number of non-zero entries in $\mathbf{w}$ increases, and the search concludes when the sparsest vector $\mathbf{w}$ that satisfies constraint \eqref{eq.constraint} has been identified.

When candidate solutions for the $L$ columns of $\hat{\mathbf{E}}_{\overbar{\mathcal{E}}}$ have been obtained and $\hat{\mathbf{E}}_{\overbar{\mathcal{E}}}$ has been evaluated, an estimate of $\mathbf{X}_{\overbar{\mathcal{E}}}$, denoted by $\hat{\mathbf{X}}_{\overbar{\mathcal{E}}}$, can be derived using \eqref{eq.PPR_Y_corrupted}:
\begin{equation}
\label{eq.XE_estimate}
\hat{\mathbf{X}}_{\overbar{\mathcal{E}}}=\mathbf{Y}_{\overbar{\mathcal{E}}}-\hat{\mathbf{E}}_{\overbar{\mathcal{E}}}.
\end{equation}
Matrix $\hat{\mathbf{X}}_{\overbar{\mathcal{E}}}$ is the outcome of the eavesdropper's attempt to repair the received matrix $\mathbf{Y}_{\overbar{\mathcal{E}}}$. Let $\nu$ denote the number of rows in $\hat{\mathbf{X}}_{\overbar{\mathcal{E}}}$ that correspond to successfully repaired coded packets, e.g., packets that passed Cyclic Redundancy Check (CRC) verification. The indices of the $\nu$ repaired coded packets are removed from set $\overbar{\mathcal{E}}$ and added to set $\mathcal{E}$, and the corresponding rows of $\hat{\mathbf{X}}_{\overbar{\mathcal{E}}}$ are moved to $\mathbf{X}_{\mathcal{E}}$. The cardinalities of sets $\overbar{\mathcal{E}}$ and $\mathcal{E}$ change to $N-N_\mathrm{E}-\nu$ and $N_\mathrm{E}+\nu$, respectively, while the dimensions of $\hat{\mathbf{X}}_{\overbar{\mathcal{E}}}$ and $\mathbf{X}_{\mathcal{E}}$ change to $(N-N_\mathrm{E}-\nu)\times L$ and $(N_\mathrm{E}+\nu)\times L$, respectively. The indices of the $\nu$ repaired coded packets are also used to identify the rows of the generator matrix $\mathbf{G}$ that should be inserted in $\mathbf{G}_{\mathcal{E}}$. If PPR increases the rank of the enlarged $(N_\mathrm{E}+\nu)\times K$ matrix $\mathbf{G}_{\mathcal{E}}$ to $K$, the RLC decoder at the eavesdropper will recover the source message.

The computational complexity that PPR adds to RLC decoding for operations in $\mathbb{F}_2$ has been discussed in \cite{Mohammadi2016}. Essentially, PPR allows the eavesdropper to improve the chances of repairing and decoding the collected coded packets at the cost of increased complexity. This fact makes PPR suitable for offline use. The probability that the eavesdropper will recover the source message, referred to as the \textit{intercept probability}~\cite{Khan2015b}, when RLC decoding is used, and the challenges in computing the intercept probability when PPR-assisted RLC decoding is employed, are discussed in the following section.


\section{Decoding and Intercept Probabilities}
\label{sec.InterceptProb}

We initially investigated the inherent secrecy of RLC in~\cite{Khan2015b} and obtained a closed-form expression for the intercept probability when both the destination and the eavesdropper employ RLC decoding. This expression will serve as a benchmark for the intercept probability when the eavesdropper uses PPR-assisted RLC decoding. For this reason, and for the sake of completeness, we summarize below the analysis in~\cite{Khan2015b}.

\subsection{RLC Decoding}
\label{sec.RLC_Dec_Analysis}

Recall that the source broadcasts $N\geq K$ coded packets, where $N$ is constrained by an upper limit $N_\mathrm{max}$. The destination and the eavesdropper receive $N_\mathrm{D}\leq N$ and $N_\mathrm{E}\leq N$ error-free coded packets, respectively. Let $R$ denote a receiver, either the destination $\mathrm{D}$ or the eavesdropper $\mathrm{E}$, that is, $R\in\{\textrm{D},\textrm{E}\}$. Also, let $X$ be a random variable, which represents the number of transmitted coded packets that are required for $R$ to recover the $K$ source packets. The Cumulative Distribution Function (CDF) of $X$, which describes the probability that the receiver $R$ will recover the $K$ source packets after $N$ coded packets have been transmitted, for $K\leq N\leq N_\mathrm{max}$, takes the form:
\begin{equation}
\label{eq:prob_cdf}
\begin{split}
F_R(N)&=\textrm{Pr}\left\{X \leq N\right\}\\
&=\sum_{N_R=K}^{N}\!\binom{N}{N_R}(1-\varepsilon_R)^{N_R}\varepsilon_{R}^{N-N_R}\;P_\mathrm{s}(N,N_R,K).
\end{split}
\end{equation}
Note that $P_\mathrm{s}(N,N_R,K)$ has been used in place of $P_\mathrm{ns}(N_R,K)$ in \eqref{eq:prob_cdf} because typical non-systematic RLC, which was considered in \cite{Khan2015b}, has been replaced by full-rank non-systematic RLC, which was defined in Section \ref{sec.RLC_encoding}. Based on the definition of $F_R(N)$, the probability that the destination $\mathrm{D}$ will recover the source message by the time $N_\mathrm{max}$ coded packets have been transmitted or sooner, is given by $F_R(N_\mathrm{max})$ for $R=\textrm{D}$, i.e.,
\begin{equation}
\label{eq:dec_prob}
P_\textrm{dec}(N_\mathrm{max}) = F_\mathrm{D}(N_\mathrm{max}),
\end{equation}
and is commonly referred to as the \textit{decoding probability}.

The Probability Mass Function (PMF) of $X$, which is the probability that the receiver $R$ will recover the $K$ source packets only after the $N$-th coded packet has been transmitted, but not sooner, can be obtained from the CDF as follows:
\begin{equation}
\begin{split}
\label{eq:prob_pmf}
f_R(N)&= \textrm{Pr}\left\{X = N\right\}\\
&=\left\{
	\begin{array}{ll}
		F_R(N)-F_R(N-1), &\!\!\!\mbox{if }K<N\leq N_\mathrm{max}\\[0.5em]
		F_R(K), &\!\!\!\mbox{if }N = K.
	\end{array}
\right.
\end{split}
\end{equation}

The intercept probability can be expressed as the sum of the following two terms:
\begin{equation}
\begin{split}
\label{eq:prob_ic_FT}
P_\textrm{int}(N_\mathrm{max}) = &\sum_{N=K}^{N_\mathrm{max}}\!f_\textrm{D}(N)\,F_\textrm{E}(N)\\
&+F_\textrm{E}(N_\mathrm{max})\bigl[\,1-F_\textrm{D}(N_\mathrm{max})\bigr],
\end{split}
\end{equation}
where $F_\textrm{E}(\cdot)$ and $F_\textrm{D}(\cdot)$ can be obtained from \eqref{eq:prob_cdf} for $R=\textrm{E}$ and $R=\textrm{D}$, respectively, and $f_\textrm{D}(\cdot)$ can be obtained from \eqref{eq:prob_pmf} for $R=\textrm{D}$. The first term in \eqref{eq:prob_ic_FT} denotes the probability that the destination will recover the source message after the $N$-th coded packet has been transmitted, for $K\leq N\leq N_\mathrm{max}$, and the eavesdropper will recover it sooner or concurrently with the destination. The second term in \eqref{eq:prob_ic_FT} represents the probability that only the eavesdropper will be successful in recovering the source message when the maximum number of coded packet transmissions, $N_\mathrm{max}$, is reached.

Expression \eqref{eq:prob_ic_FT} assumes that the sequence of coded packets delivered over the source-to-destination link is independent of the sequence delivered over the source-to-eavesdropper link. In practice, the interdependence of the two sequences becomes weaker as the value of the product $N\varepsilon_\mathrm{D}$ or $N\varepsilon_\mathrm{E}$ increases  \cite{Khan2015a} and, hence, the accuracy of expression \eqref{eq:prob_ic_FT} improves.

\subsection{PPR-assisted RLC Decoding at the Eavesdropper}

Whereas the rank of $\mathbf{G}_\mathcal{E}$ is important in RLC decoding at the eavesdropper, the \textit{spark} of the transpose of $\mathbf{H}_{\overbar{\mathcal{E}}}$ plays a key role in PPR-assisted RLC decoding. The term `spark' of a matrix was first coined by Donoho and Elad in \cite{Donoho2003} and is defined as follows for the case of $\left(\mathbf{H}_{\overbar{\mathcal{E}}}\right)^\top$:
\begin{newdef}[\cite{Elad2010}, p.23]
The spark of $\left(\mathbf{H}_{\overbar{\mathcal{E}}}\right)^\top$ is the smallest number of columns from $\left(\mathbf{H}_{\overbar{\mathcal{E}}}\right)^\top$ that are linearly dependent. Using mathematical notation, we can write:
\begin{IEEEeqnarray*}{ll}
\mathrm{spark}\left(\left(\mathbf{H}_{\overbar{\mathcal{E}}}\right)^\top\right):=&\min_{\mathbf{z}}\{\lVert\mathbf{z}\rVert_0\}\\
&\text{subject to}\; \left(\mathbf{H}_{\overbar{\mathcal{E}}}\right)^\top\!\mathbf{z}^{\top}=\mathbf{0}\;\,\text{for}\;\,\mathbf{z}\neq\mathbf{0}.
\end{IEEEeqnarray*}
\end{newdef}
If we consider the $L$ systems of equations defined in \eqref{eq.syndrome_decoding_partial_per_col} and focus on the solution to the $j$-th system described in \eqref{eq.L0_minimization}, the spark of $\left(\mathbf{H}_{\overbar{\mathcal{E}}}\right)^\top$ gives the following criterion for the uniqueness of sparse solutions:
\begin{theorem}[\cite{Elad2010}, p.24]
\label{th.uniqueness}
If a system of equations, as in \eqref{eq.constraint}, has a solution $\mathbf{w}^\top$ obeying:
\begin{equation}
\label{eq.sol_uniqueness}
\mathrm{spark}\left(\left(\mathbf{H}_{\overbar{\mathcal{E}}}\right)^\top\right)>2\lVert \mathbf{w} \rVert_0,
\end{equation}
then this solution is necessarily the sparsest possible.
\end{theorem}

Computation of the intercept probability for an eavesdropper employing online RLC decoding and offline PPR-assisted RLC decoding requires the inclusion of additional terms in~\eqref{eq:prob_ic_FT}. These terms should account for the probability that the RLC decoder at the eavesdropper will fail to recover the source message at first, but will succeed if PPR repairs a sufficient number of erroneously received coded packets. More specifically, a correctly estimated column of $\mathbf{E}_{\overbar{\mathcal{E}}}$:
\begin{enumerate}
\item either satisfies \eqref{eq.constraint} and obeys \eqref{eq.sol_uniqueness}, i.e., the estimated column is the sparsest possible solution,
\item or satisfies \eqref{eq.constraint} without obeying \eqref{eq.sol_uniqueness}.
\end{enumerate}
PPR-assisted RLC decoding will recover the $K$ source packets:
\begin{itemize}
\item if PPR correctly estimates all of the $L$ columns of $\mathbf{E}_{\overbar{\mathcal{E}}}$, because of a combination of (a) and (b), thus repairing all of the erroneously received coded packets and reconstructing the full-rank generator matrix $\mathbf{G}$, or
\item if PPR correctly estimates some of the $L$ columns of $\mathbf{E}_{\overbar{\mathcal{E}}}$, because of a combination of (a) and (b), and repairs some of the erroneous coded packets, leading to an increase in the rank of $\mathbf{G}_{\mathcal{E}}$ to $K$, as described in Section \ref{sec.PPR_at_eavesdropper}.
\end{itemize}

Derivation of the probability of fully or partially reconstructing matrix $\mathbf{E}_{\overbar{\mathcal{E}}}$ and subsequently obtaining an increased-size full-rank matrix $\mathbf{G}_{\mathcal{E}}$, where $\mathbf{G}_{\mathcal{E}}$ and $\mathbf{E}_{\overbar{\mathcal{E}}}$ are random matrices over $\mathbb{F}_q$, is a non-trivial problem and a closed-form expression is not available in the literature, to the best of our knowledge. For this reason, we use the analysis in Section~\ref{sec.RLC_Dec_Analysis} to develop and validate the simulation results of the system in Fig.~\ref{fig:system_model}, when the eavesdropper employs RLC decoding only. Then, we equip the eavesdropper with PPR capabilities and carry out simulations to measure the impact of PPR on the intercept probability, as we explain in the following section.


\section{Results and Discussion}
\label{sec.Results}

In order to investigate the inherent secrecy of RLC, we consider a source employing full-rank non-systematic RLC over $\mathbb{F}_2$ to encode a message that has been segmented into $K=20$ packets, each consisting of  $L=128$ bits. The intercept probability $P_\mathrm{int}$ has been measured for different values of packet error probabilities $\varepsilon_\mathrm{D}$ and $\varepsilon_\mathrm{E}$, when the destination employs RLC decoding and the eavesdropper also relies on RLC decoding or combines PPR with RLC decoding. Of the two PPR methods summarized in Section~\ref{sec.PPR_at_eavesdropper}, we implemented syndrome decoding (SD). The CS-based method can recover the source message with a slightly higher probability than SD for a broad range of channel conditions but at a significantly higher computational cost \cite{Mohammadi2016}. Given that the information value of the source message may decline with time, we opted for the less time-consuming SD method to complement RLC decoding. We refer to this joint scheme as `RLC with SD'.

\begin{figure}[t]
\centering
\includegraphics[width=0.99\columnwidth]{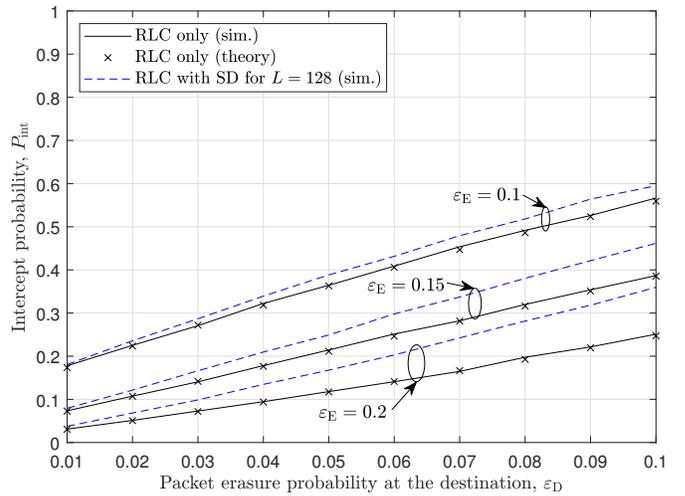}
\caption{Intercept probabilities achieved by RLC and by RLC with SD as a function of $\varepsilon_\mathrm{D}$ for $K=20$ and $\varepsilon_{\mathrm{E}}\in\{0.1,0.15,0.2\}$. The source ceases transmission when the destination recovers the source message ($P_\mathrm{dec}=1$).}
\label{fig:results_1}
\end{figure}

We first study the case where the source broadcasts coded packets until the destination recovers the source message, that is, the number of transmitted coded packets $N$ is not constrained by an upper limit $N_\mathrm{max}$. This is equivalent to setting $N_\mathrm{max}\rightarrow\infty$ in order to achieve $P_\mathrm{dec}=1$. The intercept probability has been plotted against $\varepsilon_\mathrm{D}$ in Fig.~\ref{fig:results_1}, where $\varepsilon_\mathrm{D}$ varies from $0.01$ to $0.1$ and $\varepsilon_\mathrm{E}$ takes values in~$\{0.1, 0.15, 0.2\}$. For a fixed value of $\varepsilon_\mathrm{E}$, we observe that as the value of $\varepsilon_\mathrm{D}$ increases and the source is compelled to broadcast more coded packets to ensure that the destination will recover the source message, the chances of the eavesdropper also reconstructing the source message increase. Note that expression~\eqref{eq:prob_ic_FT}, which has been adapted for full-rank non-systematic RLC encoding proposed in Section \ref{sec.RLC_encoding}, generates theoretical predictions for the intercept probability that closely match simulation results when the eavesdropper employs RLC decoding. Fig.~\ref{fig:results_1} shows that if SD is used offline to complement RLC decoding, a gain in the intercept probability will be attained, which increases with $\varepsilon_\mathrm{D}$ and, in particular, with $\varepsilon_\mathrm{E}$.

\begin{figure}[t]
\centering
\includegraphics[width=0.99\columnwidth]{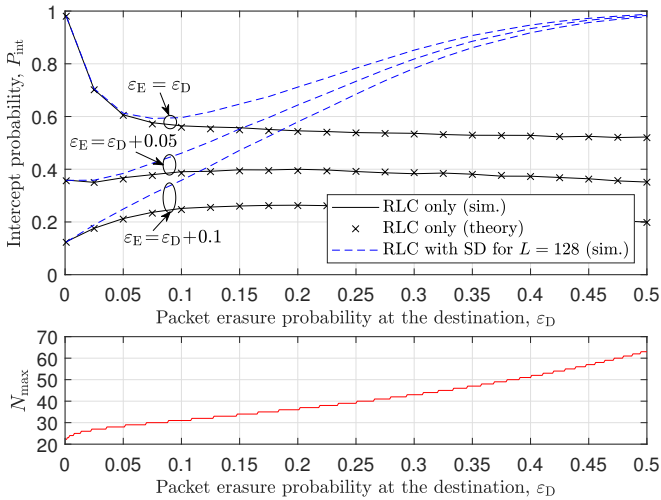}
\caption{Top: Intercept probabilities achieved by RLC and by RLC with SD as a function of $\varepsilon_\mathrm{D}$ for $\varepsilon_{\mathrm{E}}=\varepsilon_{\mathrm{D}}+\delta$, where $\delta\in\{0,0.05,0.1\}$, $K=20$ and a varying $N_\mathrm{max}$. Bottom: For each value of $\varepsilon_\mathrm{D}$, the value of $N_\mathrm{max}$ ensures that the destination will recover the source message with probability $P_\mathrm{dec}\geq 0.99$.}
\label{fig:results_2}
\end{figure}

Let us now consider the case where the source sets a limit $N_\mathrm{max}$ on the number of transmitted coded packets, so that the decoding probability at the destination is at least $0.99$, that is, $P_\mathrm{dec}(N_\mathrm{max})\geq 0.99$ but $P_\mathrm{dec}(N_\mathrm{max}-1)<0.99$. Expression \eqref{eq:dec_prob} was used to determine $N_\mathrm{max}$ for different values of $\varepsilon_\mathrm{D}$, and their relationship has been plotted in Fig.~\ref{fig:results_2} (bottom). The intercept probability for increasing values of $\varepsilon_\mathrm{D}$ and $N_\mathrm{max}$ is shown in Fig.~\ref{fig:results_2} (top), where $\varepsilon_\mathrm{E}=\varepsilon_\mathrm{D}+\delta$ for $\delta\in\{0, 0.05, 0.1\}$. We observe that, as channel conditions deteriorate, the \mbox{inherent} secrecy of RLC is not severely compromised when the eavesdropper relies on RLC decoding to obtain the source message; for $\varepsilon_\mathrm{E}>\varepsilon_\mathrm{D}$, the intercept probability increases marginally but then plateaus and reduces slowly, as $\varepsilon_\mathrm{D}$ shifts from~$0$~to~$0.5$. However, when the eavesdropper combines RLC with SD, the intercept probability increases noticeably and approaches $1$ for large values of $\varepsilon_\mathrm{D}$. This result establishes that a poor source-to-destination channel, which could have been intentionally degraded by an eavesdropper who generates artificial noise, greatly improves the decoding capability of the eavesdropper when RLC with SD is used, even if the eavesdropper experiences similar or worse channel conditions than the destination.

\begin{figure}[t]
\centering
\includegraphics[width=0.99\columnwidth]{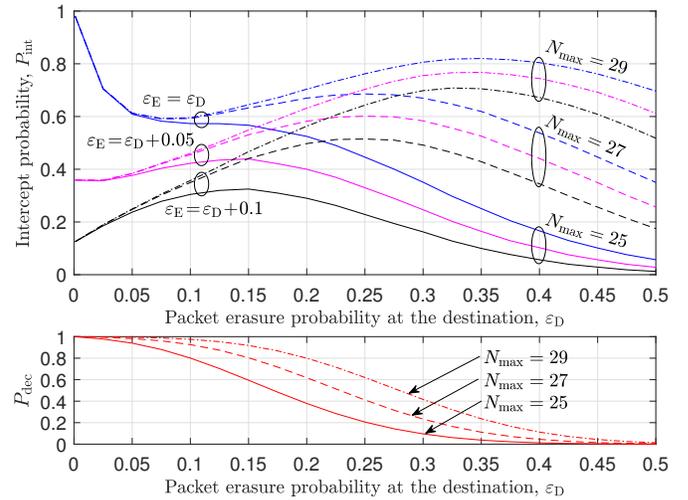}
\caption{Top: Intercept probabilities achieved by RLC with SD as a function of $\varepsilon_\mathrm{D}$ for $\varepsilon_{\mathrm{E}}=\varepsilon_{\mathrm{D}}+\delta$, where $\delta\in\{0,0.05,0.1\}$, $K=20$ and fixed values of $N_\mathrm{max}$. Bottom: Probability that the destination will recover the source message for $N_\mathrm{max}\in\{25,27,29\}$.}
\label{fig:results_3}
\end{figure}

Given that a scenario where $\varepsilon_\mathrm{E}=\varepsilon_\mathrm{D}$ for $\varepsilon_\mathrm{D}\rightarrow 0$ is not likely to occur, limiting the value of the intercept probability for all other cases is of importance for systems that rely on information theoretic security. Instead of raising the value of $N_\mathrm{max}$ as $\varepsilon_\mathrm{D}$ increases in order to maintain a fixed decoding probability at the destination, as in Fig.~\ref{fig:results_2}, we fix the value of $N_\mathrm{max}$ in Fig.~\ref{fig:results_3} and trade reliability for secrecy. We observe that a decrease in the value of $N_\mathrm{max}$ suppresses information leakage and reduces the intercept probability, when the eavesdropper combines SD with RLC decoding, but inevitably lowers the decoding probability. The value of $N_\mathrm{max}$ also determines the range of $\varepsilon_\mathrm{D}$ values over which the system could operate. For example, if $N_\mathrm{max}=25$, the destination will achieve $P_\mathrm{dec}=0.8$ for $\varepsilon_\mathrm{D}\!=\!0.1$ while the worst-case intercept probability ($\varepsilon_\mathrm{E}\!=\!\varepsilon_\mathrm{D}$) will be limited to $P_\mathrm{int}=0.57$. However, the system may be required to support the same decoding probability \mbox{($P_\mathrm{dec}=0.8$)} for $\varepsilon_\mathrm{D}=0.2$; Fig.~\ref{fig:results_3} indicates that $N_\mathrm{max}$ will have to increase from $25$ to $29$ as long as the subsequent increase in the worst-case intercept probability ($P_\mathrm{int}=0.7$) is tolerable, e.g., because additional security measures have been implemented or the source message does not carry sensitive information.


\section{Conclusions}
\label{sec.Conclusions}

This paper investigated the impact of partial packet recovery on the inherent secrecy of random linear coding. Results demonstrated that syndrome decoding combined with random linear decoding can boost the probability of an eavesdropper recovering a message transmitted from a source to a legitimate destination. Of particular interest was the observation that this probability increases as the channel conditions between the source and the destination, but also between the source and the eavesdropper, deteriorate. The eavesdropper has thus an incentive to introduce artificial noise and actively impair data packets transmitted by the source. If information theoretic security is essential, system requirements should dictate a limit on the number of packet transmissions and a threshold on the channel conditions, beyond which transmission should cease in order to keep low the probability of the eavesdropper reconstructing the source message.

\section*{Acknowledgement and Research Reproducibility}
\label{sec.ack}

This work was motivated by the outcomes of the third-year project completed by Adam Matthews and supervised by the author. The MATLAB\textsuperscript{\textregistered} code for the simulations and the generation of the figures in Section~\ref{sec.Results} can be downloaded from \url{https://github.com/IoannisChatzigeorgiou/VTC2022-Spring}.


\bibliographystyle{IEEEtran}
\bibliography{IEEEabrv,references}


\end{document}